# A New Physically Triggered Cell Death via Transbarrier Contactless Cold Atmospheric Plasma Treatment of Cancer Cells

Dayun Yan [1,*,§], Qihui Wang [1,§], Manish Adhikari [1,§], Alisa Malyavko[2], Li Lin [1], Denis B. Zolotukhin [1], Xiaoliang Yao[1], Megan Kirschner [1], Jonathan H. Sherman [3], Michael Keidar [1,*].

[1]. Department of Mechanical and Aerospace Engineering, The George Washington University, Science & Engineering Hall, 800 22nd Street, NW, Washington, DC, 20052, USA.

[2]. School of Medicine and Health Sciences, The George Washington University, 2300 I St NW, Washington, DC, 20052, USA.

[3]. Neurological Surgery, The George Washington University, Foggy Bottom South Pavilion, 22nd Street, NW, 7th Floor, Washington, DC, 20037, USA.

[§] These authors contributed equally to this manuscript.

* Corresponding authors: Dayun Yan, ydy2012@gwmail.gwu.edu, 202-994-6929.

  Michael Keidar, keidar@gwu.edu, 202-994-6929.

**Abstract:** For years, extensive efforts have been made to discover effective, non-invasive anti-cancer therapies. Cold atmospheric plasma (CAP), is a near room temperature ionized gas composed of reactive species, charged particles, neutral particles, and electrons. CAP also has several physical factors including thermal radiation, ultraviolet (UV) radiation, and electromagnetic (EM) waves. Most of the previously reported biological effects of CAP have relied on direct contact between bulk plasma and cells, resulting in the chemical effects generally seen after CAP treatment. In this paper, we demonstrate that the electromagnetic emission produced by CAP can lead to the death of B16F10 melanoma cancer cells via a transbarrier contactless method. When compared with the effect of reactive species, the effect of the physical factors causes much greater growth inhibition. The physical-triggered growth inhibition is due to a new type of cell death, characterized by rapid leakage of bulk water from the cells, resulting in bubbles on the cell membrane, and cytoplasm shrinkage. The results of this study introduce a new possible mechanism of CAP induced cancer cell death and build a foundation for CAP to be used as a non-invasive cancer treatment in the future.

**Introduction.**

Skin cancer is one of the most common global malignancies. This broad category can be divided into melanoma and non-melanoma skin cancer. Although non-melanoma cancers such as basal cell carcinomas and squamous cell carcinomas are more common in the population, melanoma is the deadliest category of skin cancers [1–3]. Currently, there is a variety of different treatment options for melanoma, however, there is a constant need for more effective and non-invasive therapies [4–6].

CAP is a near room temperature ionized gas achieved in a non-equilibrium discharge at atmospheric pressure in the air mixed with inert gases such as helium and argon [7,8]. CAP is composed of non-equilibrium ionized particles, including neutral particles, ions, and electrons as well as highly reactive species and other chemical factors. The reactive species impact a variety of cells such as bacterial, plant, and mammalian cells in a multitude of ways [9,10]. However, a direct interaction of reactive species and cells is necessary for the biological and chemical effects of CAP. Currently, the conventional method of using CAP is based on an invasive method which limits its range of application.

The conventional chemical CAP treatment leads to apoptosis of mammalian cells, a natural cellular response to an increase in reactive species intracellularly. CAP triggered apoptosis follows a typical apoptotic pathway with changes in cell morphology including cell shrinkage, DNA fragmentation, membrane budding, and formation of the final apoptotic body with membranous vesicles [11,12]. Although some groups have found markers of necrosis based on flow cytometry data, the corresponding mechanism is not clear or well understood making it difficult to definitively conclude necrotic cell death.

Aside from chemical factors, CAP produces three physical factors including thermal radiation, ultraviolet (UV) radiation, and electromagnetic (EM) waves in the range of 10-100 GHz [13,14]. Until now, all these physical factors have been thought to have negligible roles in cancer cell function and death following CAP treatment due to a lack of direct evidence of cellular responses to these factors [15,16]. However, the physical factors may be responsible or play important roles in exerting non-invasive biological effects on cancer cells during CAP treatment.

In this study, we show the experimental evidence of a strong anti-cancer effect on B16F10 melanoma cells *in vitro* caused by electromagnetic emission from CAP. We initially demonstrate that CAP treatment is able to cause a great amount of cell death through an indirect, contactless method across a physical barrier (> 1 mm). When cell viability is compared after conventional chemical treatment and this novel physical treatment, physical treatment achieves a more significant reduction in cell viability. A new type of cell death has also been noted in cells after physical CAP treatment which could be contributing to the physically triggered anti-cancer effect. This cell death is characterized by rapid leakage of bulk water from the cytoplasmic membrane which results in visible bubbling on the cellular surface. The bubbling mechanism has been investigated through exerting osmotic pressure using hypotonic solutions. This study builds the foundation to understand the roles of physical factors of CAP in cancer cell death. With this knowledge, a more complete mechanism of CAP can be proposed which will allow for greater translational use of this novel, non-invasive anti-cancer therapy.

**Results**

**Effects of chemical and physical CAP treatment.**

Conventionally, *in vitro* CAP treatment is performed by treating cells cultured in dishes or multi-well plates where the plasma directly interacts with the medium covering the cells (Figure 1a). This strategy is an effective way to study the effect of reactive species on mammalian cells, particularly the cancer cells in this case. However, this strategy may interfere with or completely block the physical effect of CAP on cells due to the bulk aqueous environment. Conventional CAP treatment also results in both the chemical and physical factors affecting the cell making it difficult to unravel the effect of each factor and obtain a clear understanding of the underlying mechanism. To observe the purely physical effect of CAP, we used a novel technique to completely preclude all chemical factors. Throughout the study, physical CAP treatment was performed by inverting multi-well plates such as 96-well and 12-well plates and treating from the back as shown in Figure 1b. The same protocol was done for various dish sizes ranging from 100 mm to 35 mm dishes. The detailed description of the protocols for both chemical and physical CAP treatment is stated in the Methods and Materials. Physical CAP treatment causes 2D growth inhibition which is shown on treated 96-well plates using 2D cell viability maps. The protocol to generate 2D cell viability maps is illustrated in Figure S1. In addition, we demonstrate that simply keeping the plates in an inverted position will not cause growth inhibition of B16F10 cells (Figure S2).

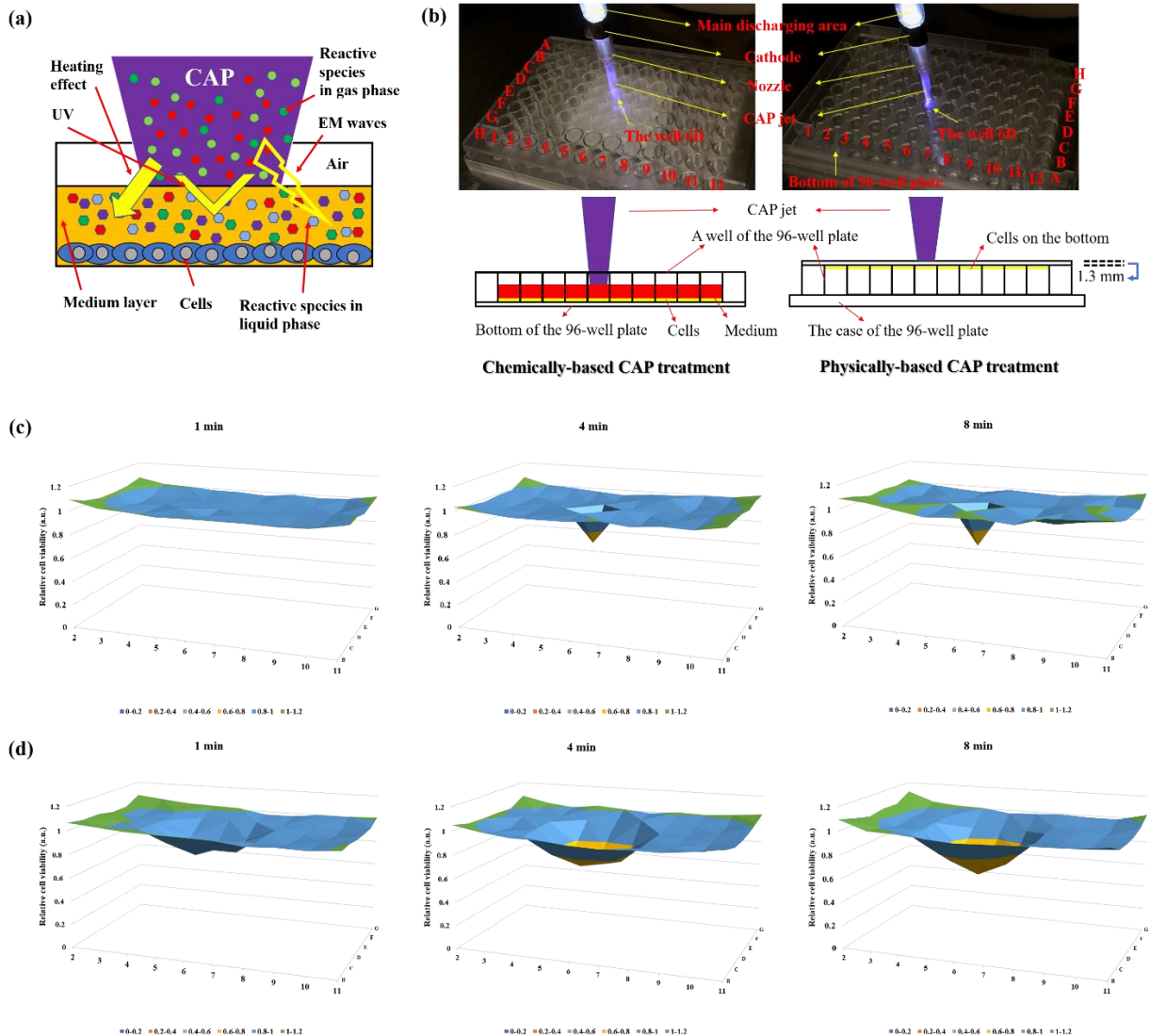

Figure 1. The chemically based CAP treatment and the physically based CAP treatment. (a) The chemical and physical factors in the CAP treatment. (b) Chemical treatment and physical treatment. Here, as an example, the physical CAP treatment was performed by treating the bottom of an inverted 96-well plate. The treated well was the well 6D on a 96-well plate. The gap between the CAP nozzle and the bottom of the 96-well plate was 25 mm. (c) The 2D cell viability maps of the chemically based CAP treatment on a 96-well plate. (d) The 2D cell viability maps of the physically based CAP treatment on a 96-well plate. All experimental conditions were the same in 2 different treatment strategies. The data are presented as the mean of 4 independent experiments. The original data are shown in Figure S3.

The conventional efficacy of chemical CAP treatment is determined by the reactive species. Therefore, if a cell line is resistant to reactive oxygen species (ROS) or reactive nitrogen species (RNS), it will also be resistant to chemical CAP treatment. B16F10 melanoma cells are very resistant to reactive species, particularly ROS such as $H_2O_2$ [17]. We initially compared the amount of growth inhibition seen following chemical CAP treatment versus physical CAP treatment. Compared to chemical CAP treatment, physical CAP treatment not only has a greater impact on growth inhibition but is also able to impact a wider area (Figure 1c and Figure 1d). Chemical CAP treatment only causes growth inhibition in the treated well, 6D, whereas physical CAP treatment noticeably inhibits the growth of melanoma cells in at least eight wells surrounding the treated well 6D.

The CAP treatment on a 12-well plate shows a similar but more drastic difference between chemical CAP and physical CAP treatment. Protocols are illustrated in the Methods and Materials. It is found that 3 minutes of chemical CAP treatment causes very minimal growth inhibition of B16F10 cells. On the other hand, just 1.5 min of physical CAP treatment resulted in nearly 60% growth inhibition of B16F10 cells (Figure 2a, Figure 2b, and Figure 2c). The increased growth inhibition following physical CAP treatment is thought to be a result of a new type of cell death. Microscopic imagining was performed at one day and two days post-treatment and novel changes were seen in these physically CAP treated cells (Figure 2d and Figure 2e). These cell changes can be characterized by aggregation of the cytoplasm toward the nucleus and the nucleus becoming colorless, which can signify a loss of nuclear components. We noted that these changes remained for several days following the treatment. We did not witness any cell duplication, cell movement, or cell to cell interaction leading us to believe that these cells were in a somewhat "fixed" state. We conclude that these new cellular changes are evidence of a new type of cell death, which is independent of chemical CAP factors and is the reason for physical CAP treatment leading to such strong growth inhibition on a reactive species resistant cell line B16F10.

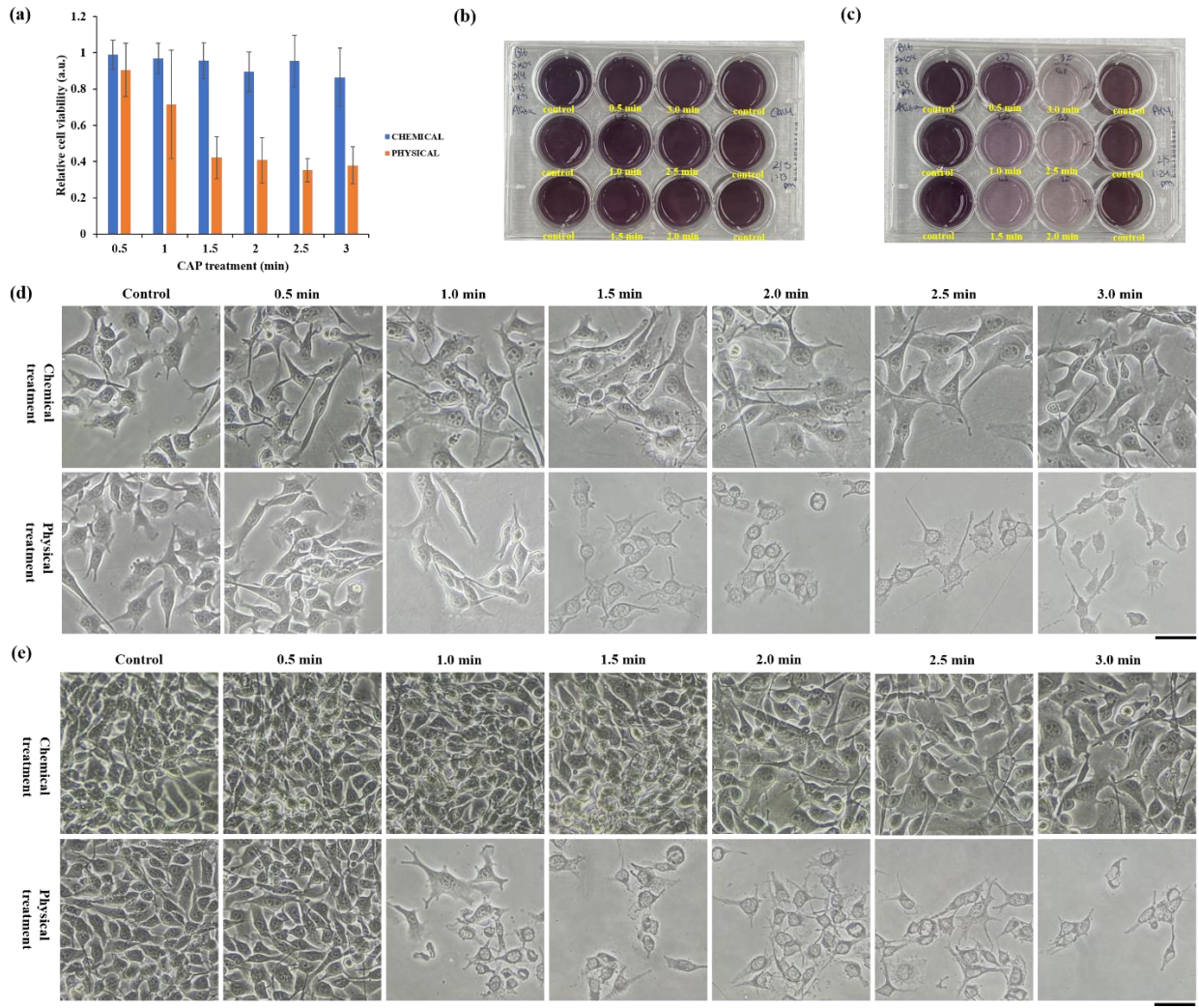

Figure 2. The drastically different cellular responses to the chemically and physically based CAP treatment. (a) The chemical and physical inhibition of the cell viability. The data are presented as the mean of 6 independent experiments. (b) The MTT assay on a 12-well plate after the chemically based treatment. (c) The MTT assay on a 12-well plate after the physically based treatment. (d) The microscopic images of the cells one day after the treatment (0.5 min ~3 min). (e) The microscopic images of the cells two days after the treatment (0.5 min ~3 min). The scale bar is 50 µm (black). For the chemical treatment, the gap between the nozzle to the cells was 37 mm. For the physical treatment, the gap between the nozzle to the bottom surface was 19 mm. The flow rate was 1.53 lpm. The microscopic imaging was performed using a Nikon TS100 inverted phase contrast microscope.

**Bubbling is an early feature of the novel cell death.**

Apoptosis is the main mechanism of cell death seen in nearly all *in vitro* studies, however, we have observed a new type of cell death with novel corresponding cellular changes. To further investigate the transition from a normal cellular shape to the observed "fixed" cellular shape, we started by observing the immediate morphological changes of B16F10 cells after physical CAP treatment. We did not do in-situ observation during the CAP treatment. However, we did observe the changes occurring in B16F10 cells immediately after a CAP treatment length of 1 min, 2 min, and 3 min. As shown in Figure 3a, a noticeable change in cell morphology is initially noted after a CAP treatment of 2 min. A significant shrinkage of the cytoplasm and a change in the nucleus can be observed. Several bubbles can be seen forming on the cellular membrane of CAP treated cells as well, some of which are as large as the cell they are originating from. When the CAP treatment extends to 3 min, the bubbles disappear and the nucleus adapts a checkboard color pattern, similar to the "fixed" cells shown in Figure 2d and Figure 2e. At this time point, the nucleus is observed to have more white areas compared to observations made at shorter treatment time points. Additionally, the black dots, which could be the nucleoli, become more apparent in treated cells due to the whitening of the nuclei compared to cells in the control group. The size of the nucleus does not show any noticeable change. This cell morphology is a typical feature of physically treated cells immediately after treatment, which directly explains the "fixed" cell features observed 1 and 2 days after CAP treatment (Figure 2d and Figure 2e).

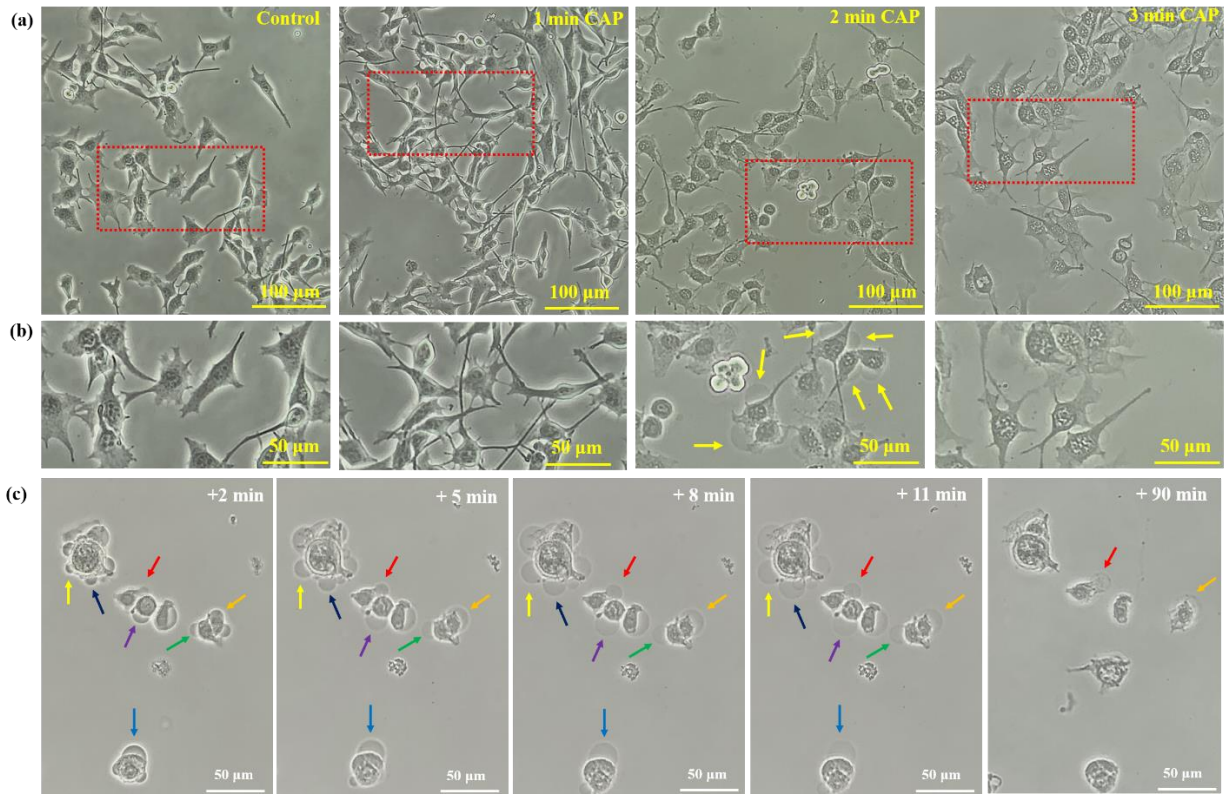

Figure 3. The quick morphological change of B16F10 cells immediately after the CAP treatment. (a) Imaging with a low magnification. (b) The zoomed in images for the red dotted boxes in (a). The blue arrows mark the bubbles on the cellular membranes. B16F10 cells were seeded with a density of 1 x $10^5$ cells/mL on a 35-mm glass bottom dish and cultured for one day before the treatment. After the treatment, 1.5 mL medium was added in the 35-mm glass bottom dish immediately (<30 s). The images were taken 8 min after the treatment. (c) The growth of single bubbles after a 2 min of CAP treatment. For all cases, the gap between the nozzle to the bottom surface was 27 mm. The flow rate was 1.53 lpm. The arrows mark the growth of 7 specific bubbles on the cellular membrane. '+ x min' means the photo was taken at x min after the CAP treatment. In the right most photo, the disappearance of certain cells and appearance of new cells could be due to the detachment and the migration of cells, respectively. The images were taken by a Nikon TS100 inverted phase contrast microscope.

The bubbling on the cell membrane is the most noticeable feature of the physical effect on B16F10 cells. Typical bubbling seen in B16F10 cells is presented in Figure S4. We performed a continuous observation of the bubbling over a while after physical CAP treatment. To observe clear bubble growth on a single cell, the seeding cell density should be relatively low (4 x $10^4$ cells/mL). The cells were seeded in 35 mm glass-bottom dishes and cultured for one day before treatment. The growth of a bubble typically takes 8-11 min after CAP treatment (Figure 3b). As the bubbles grow larger, the inner composition of the bubbles becomes lighter and more transparent. This could suggest a slow dilution of the internal components as the bubble expands. No further morphological change is seen in the shrunken cells, other than what has been described previously, during this process or at the time when the bubbles reach their maximum sizes. Therefore, bubbling appears to be a process that occurs after the aggregation of the cytoplasm. The bubbles disappear from the membrane either by detachment or another mechanism approximately 90 minutes after CAP treatment. In Figure 3b, the debris of two bubbles can be seen remaining on two cells at the 90-minute after the treatment. As shown in Figure S5, we also observed the whole 30 min of evolution of the CAP-treated B16F10 cells with a higher density (1 x $10^5$ cells/mL). There is no further morphological change of the shrunken cells during the whole observation period. Many small, detached bubbles are found in the extracellular space in Figure S5, which is the evidence supporting the final detachment mechanism of the bubbles.

The phase-contrast imaging solely provides evidence of the morphological changes of cells. To investigate changes occurring at the cellular chemistry level, we performed live-cell fluorescent imaging. The detailed protocol is listed in Methods and Materials. Microtubules and DNA were stained using BioTracker 488 Green Microtubule Cytoskeleton Dye (Sigma-Aldrich) and Hoechst 33342 (ThermoFisher Scientific), respectively. The stained B16F10 cells were incubated for 15 minutes inside the $CO_2$ incubator. Then the B16F10 underwent 4 minutes of physical CAP treatment, centered on the bottom of the dish. The gap between the nozzle to the bottom surface was 27 mm and the flow rate was 1.53 lpm. After treatment, the dish was immediately placed upright and replenished with fresh RPMI-1640 media for confocal imaging. The staining was visualized and imaged with a Carl Zeiss 710 spectral confocal microscope. Due to the operation of the imaging process, the first image was captured at 20 min after CAP treatment. Therefore, we were unable to capture the immediate changes occurring in the cells before the 20-min time mark. However, as we demonstrated above, the basic features of bubbling and the cytosol shrinkage will not change since the $8^{th}$ min after the treatment.

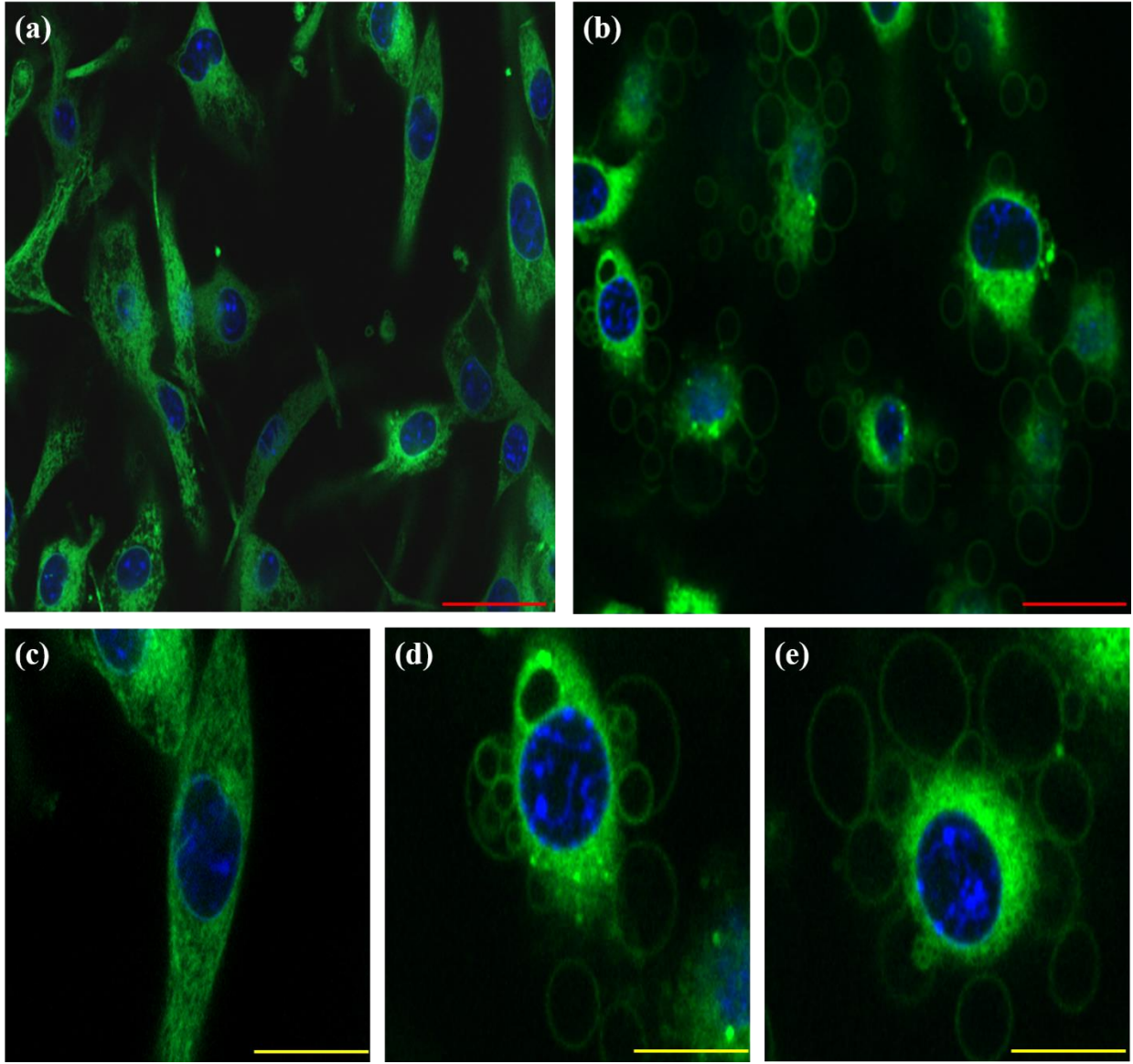

Figure 4. The live fluorescent imaging for the cellular change at 20 min after the physically based CAP treatment. (a) The control before the treatment. (b) 20 min after the treatment. (c) The zoomed in photo of the control. (d) The zoomed in photo shows the formation of bubbles on the cellular membrane. The growth of the small bubbles is illustrated in video 1. (e) The zoomed in photo of the bubbles sounding the CAP-treated cells. The microtubules and DNA (nucleus) are shown in green and blue, respectively. The scale bars are 40 μm (red) and 20 μm (yellow). All images were processed using ImageJ software.

The live fluorescent imaging confirms that the initial feature of the novel cell death is characterized by the formation of bubbles on the cellular membrane (Figure 4a and Figure 4b). The light green surface of the bubbles is most likely due to the reflection of green fluorescence on the interface of the bubbles. The lack of fluorescence within the bubbles suggests that there may be an absence of organelles in the bubbles (Figure 4c and Figure 4d). Thus, we propose that the bubbling is due to the leakage of water or cellular solution outside the cell membrane. The bubbles range in size from being as large at the nucleus to being as large as the entire cell itself (Figure 4c and Figure 4d). A "grape-like" aggregation of bubbles can be seen in Figure 4c and Figure 4d. It is believed that this aggregation pattern is due to the formation of new, smaller bubbles near or in the already formed larger bubbles. We also captured the dynamics of bubble growth on the cell membrane (video 1 and 2). In these two videos, the formation of new bubbles initiates from a single site on the cell membrane, potentially where a membrane pore or channel is present. Another clear feature is the change in the cytoskeletal structure. Filamentous microtubules are observed throughout the control cell (Figure 4c), but these filamentous features are lost in the CAP treated cells. The cytoskeleton shrinks and aggregates around the nucleus after CAP treatment (Figure 4d and Figure 4e). The clear staining of the nucleus suggests that DNA may not suffer noticeable damage immediately after CAP treatment.

The evolution of the treated cells was further investigated on a longer time scale. During this longer observation period, a gradual loss of DNA in the nucleus was seen. The rest of the cell appears to maintain its "fixed" state as seen in Figure 2d and Figure 2e without any further changes during the long period. The bubbles cannot be seen on the cells two hours after CAP treatment (Figure 5a) which is consistent with the results seen with phase contrast imagining shown in Figure 2 and

Figure 3b. The nucleus of treated cells does expand when compared to the control during the initial two hours (Figure 5b). However, at the two-hour time mark, the nucleus appears to reach a maximum size and experiences a slight decrease in size after that. The blue DNA staining in the nucleus is seen one day after treatment (Figure 5c and Figure 5d), however, it becomes very weak two days after treatment. There is no obvious blue staining of DNA in the nucleus on the third day after CAP treatment, suggesting a total loss of DNA in the nucleus at that time point. The green microtubule staining also gradually weakens by the second day after treatment. 5 days after physical CAP treatment, the final appearance of B16F10 cells resembles an empty, circular shell composed of microtubules and other potential cytoskeletal and related cytosolic components not stained for in this study (Figure 5d). This matches well with the basic feature of the 'fixed' cells seen in Figure 2d and Figure 2e. After the loss of all DNA, the cells were determined to be dead and by the sixth day after CAP treatment, we were unable to find any remnants of the cells or any evidence of green microtubule staining. These cell death features are vastly different from the typical apoptotic cell death features observed in previous studies and have not been previously reported in studies of the effect of CAP treatment on cells.

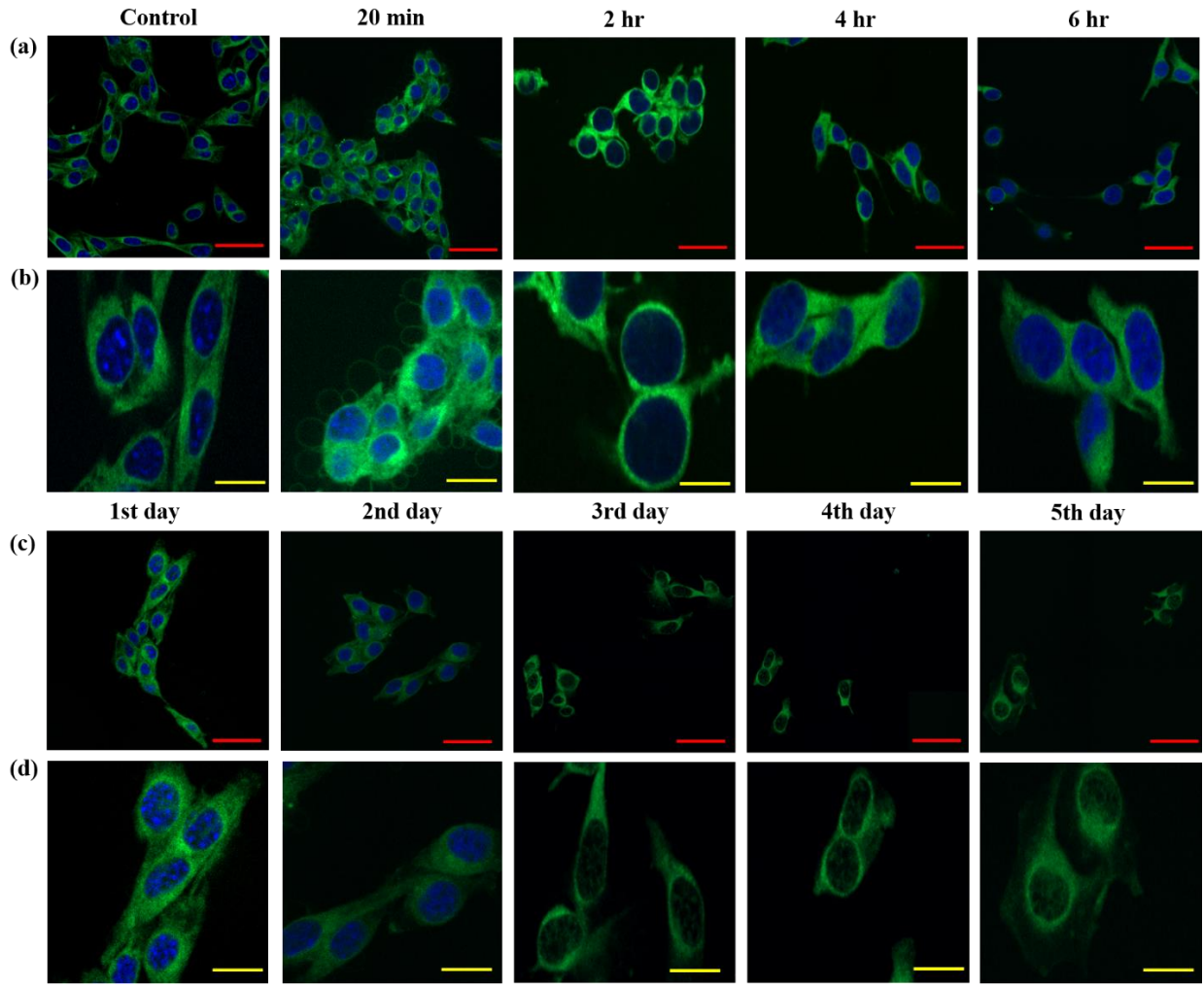

Figure 5. The fluorescent imaging of the physically based CAP-treated B16F10 cells over a long timescale. (a) The control before the treatment and the treated cells were observed 20 min, 2 hr, 4 hr, and 6 hr after the treatment. (b) The zoomed in photos. (c) The CAP-treated cells were observed 1 day, 2 days, 3 days, 4 days, and 5 days after the treatment. (d) The zoomed in photos. The microtubules and DNA (nucleus) are shown in green and blue, respectively. The gap between the nozzle to the bottom surface was 27 mm. The flow rate was 1.53 lpm. The scale bars are 40 µm (red) and 20 µm (yellow). All images were processed using ImageJ software.

The bubbling may be due to changes in osmotic pressure intracellularly. According to the common understanding of bubbling on the cell membrane, the positive intracellular osmotic pressure across the cell membrane should play an important role in regulating the bubbling effect. Such a change in osmotic pressure is mainly due to chemical factors such as the presence of a hypertonic or hypotonic solution in the extracellular chemical environment. In our study, physical treatment is not likely to change the extracellular environment. Our findings may be due to the aggregation of the cytoplasm which creates intracellular pressure which in turn can trigger the release of water or cytosolic solution across the cell membrane. Physical CAP treatment may also inflict physical damage on the cell resulting in the formation of pores on the cell membrane through which water or cytosolic solution can be released to form the bubbles.

Deionized (DI) water is a hypotonic solution that causes the flow of water into cells and can therefore potentially counteract the intracellular osmotic pressure and inhibit bubbling. According to this rationale, we prepared a series of RPMI/Milli-Q water (DI water) mixed solutions to investigate the bubbling mechanism. After 3 min of physical CAP treatment, B16F10 cells were immediately turned upright and replenished with different RPMI/Milli-Q mixed solutions. As shown in Figure 6, the density of bubbles forming from the treated cells decreases as the volume ratio of Milli-Q water increases. When the treated cells are immersed in 10 mL of Milli-Q water the bubbling is completely inhibited. Due to the influx of extracellular water into the cells, the CAP treated cells also expand in the Milli-Q water. In short, we conclude that the bubbling could be due to the intracellular osmotic pressure resulting from physical CAP treatment, however, the mechanism is still unknown.

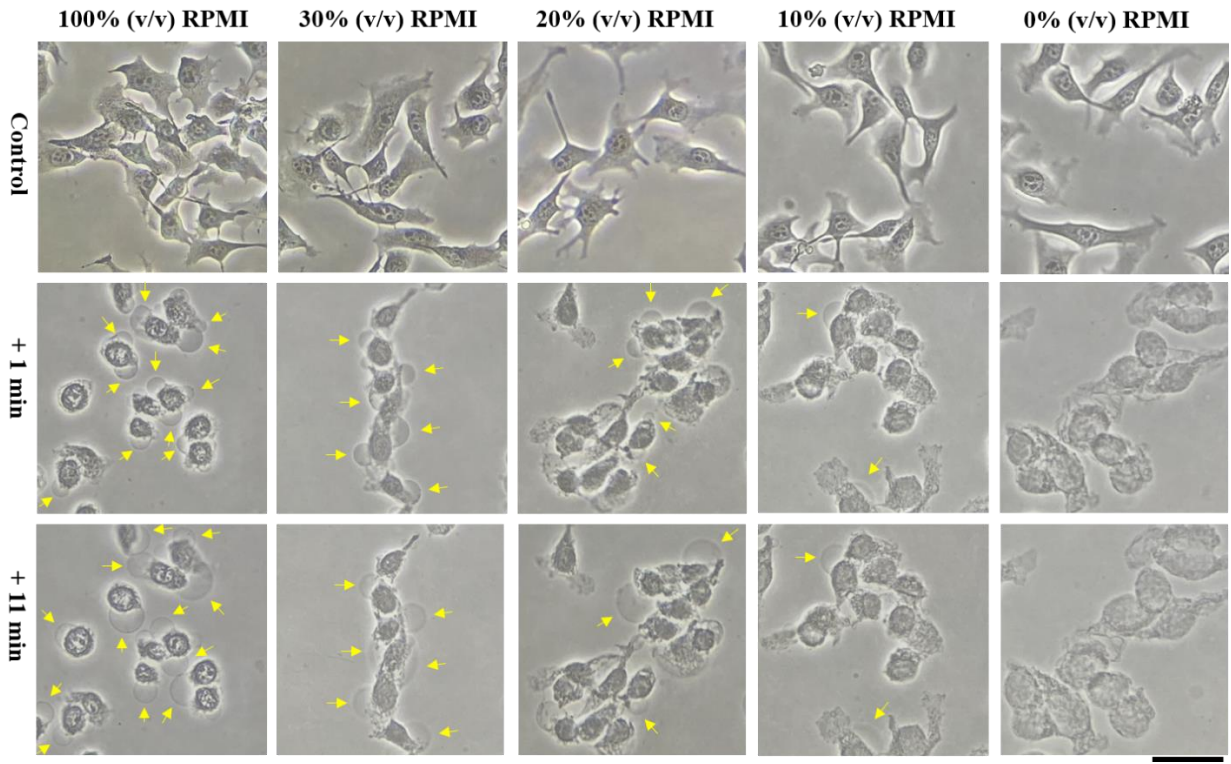

Figure 6. The hypotonic solutions can inhibit the bubbling on the physically treated melanoma cells. 10 mL cell solution was cultured in 100 mm dish with a density of 7.5 x $10^4$ cells/mL. for one day before the treatment. In each case, the CAP treatment lasted 3 min. After that, the cells were immediately (<30 s) immersed in the 100 mL of RPMI1640/Milli-Q water mixed solutions. The volume ratio % (v/v) of RPMI in the solutions was 100%, 30%, 20%, 10%, and 0%, respectively. '+ x min' means the photo was taken x min after the CAP treatment. Because the initial bubbles were not clear, only the bubbles after a 11 min of growth were marked by yellow arrows. The scale bar was 50 µm (black). The photos of control and the experimental group were taken at different places on the dish. The photos of the experimental were taken in situ. The photo of CAP jet in the treatment was show in Figure S6.

**Bubbling occurs even there is an air gap between the jet and the target.**

In all of the cases presented above, the CAP jet was directly touching the bottom of the 96 well plates, 12-well plates, 35 mm culture dishes, or 100 mm culture dishes. As emphasized previously, this treatment method does not involve direct contact of CAP with the cells or the cell medium. Here we demonstrate that the CAP jet doesn't have to directly touch the back of the culture dish to cause the effects and cell death seen after physical CAP treatment. We investigated the bubbling triggered after physical CAP treatment at different distances of the nozzle of the CAP source to the back surface of dish. Some of our conditions in this experiment created a gap between the tip of jet and the back surface of the dish as seen in Figure 7. The bubbling will not occur if the gap is too small or too large. Even when the CAP jet noticeable touches the back surface of dish, the physical effect will not cause a cellular change. A large distance between the CAP jet and the back surface of the plate of the dish will also not cause the physical effects seen previously. A gap with a distance between 27.0 mm, where the CAP jet is slightly touching the back surface, and 30.0 mm, where there is a small air gap between the tip of jet and the surface, will also trigger bubbling. In the case of a gap of 30.0mm, the distance between the tip of jet and back surface is established at 8.2 mm in the image. The corresponding distances for cases of a gap of 33.5 mm and a gap of 39.5 mm are 8.3 mm and 9.5 mm respectively. Therefore, there is a significant change in the physical effect on B16F10 cells when the distance between the tip of jet and the back surface is larger than 8.2 mm.

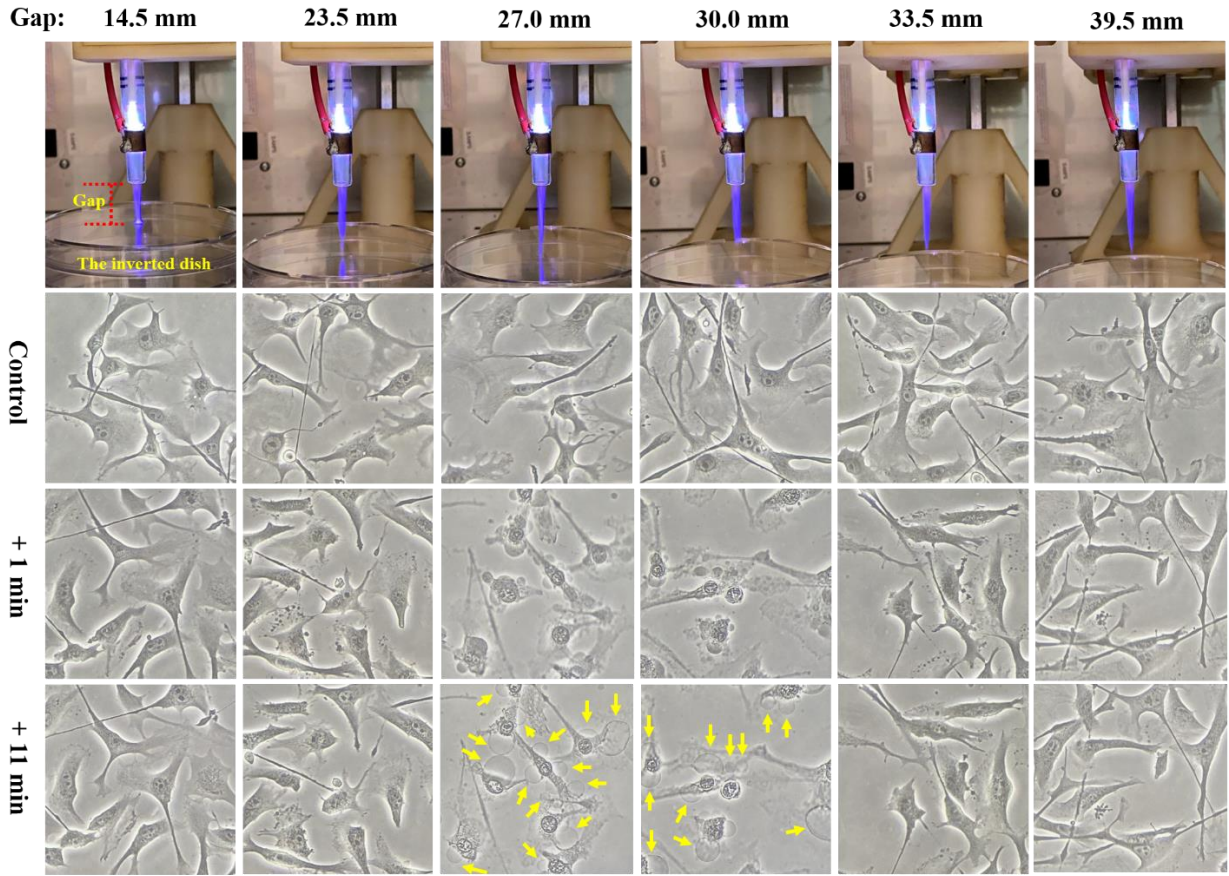

Figure 7. The bubbling only occurs when the CAP source and the bottom surface of an inverted dish has a moderate gap. Neither too small nor too long gap will trigger the bubbling. 10 mL cell solution was cultured in 100 mm dish with a density of $7.5 \times 10^4$ cells/mL. for one day before the treatment. The 100 mm dish was inverted during the treatment. The gap varied from 14.5 mm to 39.5 mm. The flow rate was 1.53 lpm for all cases. After the treatment, the cells were immediately (<30 s) immersed in 10 mL of RPMI for the microscopic imaging. Because the initial bubbles were not clear, only the bubbles after a 11 min of growth were marked by yellow arrows. The scale bar was 50 μm (black). The photos of control and the experimental group were taken at different places on the dish. The photos of the experimental were taken in situ. '+ x min' means the photo was taken at x min after the CAP treatment.

**The physical factors to cause cell death.**

To understand the specific role of individual physical factors, we began by analyzing the temperature component of the CAP. We first measured the temperature of the bottom of a 35mm glass bottom dish after 2 min of CAP treatment. The maximum temperature of the CAP-treated area was 38.5°C ± 0.4°C (Figure S7). To investigate this heating effect, B16F10 cells in a 35-mm glass-bottom dish were partially immersed in a water bath for 2 min so that the bottom of the dish was immersed in water with a temperature of 41°C ~ 43°C. The experimental protocols are illustrated in Figure S8. To simulate physical CAP treatment, the media was removed before the heating experiment. This water bath experiment did not cause any change in cell morphology when compared to the control group. In contrast, 2 min of physical CAP treatment causes the drastic cell morphology changes and cell death of B16F10 cells described earlier (Figure S8). Therefore, the heating effect alone does not cause the anti-cancer effect seen after the physical CAP treatment.

We further investigated the roles of UV radiation and EM waves emitted during physical CAP treatment. A UV reflection film was used to block the UV radiation generated by the CAP jet. The UV reflection film was purchased from VWR (Adhesive White Light-Reflecting Films, 89087-696). The characterization of the UV reflection film was shown in Figure S9. The UV reflection film has an excellent attenuation of the UV between 191 nm and 400 nm. The film was attached to the bottom surface of 96-well plates, centered on well 6D (Figure 8a). The film was removed after treatment. Two sizes of the UV film were used (5 x 5 wells, 10 x 6 wells), which covered 5 x 5 wells and 10 x 6 wells on the bottom of the 96-well plate, respectively. Similarly, to investigate the EM effect, we used a copper sheet as an EM wave blocker with two sizes (5 x 5 wells, 10 x 6 wells). The copper sheets were purchased from McMaster-CARR (9709k704). The copper sheet was set between the CAP jet and the bottom of the 96-well plate during the treatment (Figure 8b).

The sheet was also centered at the well 6D. It is found that 8 minutes of CAP treatment still causes strong growth inhibition even in the presence of a UV reflection film, regardless of size (Figure 8c and Figure 8e). Unlike the UV reflection films, the 5 x 5 well size copper sheet nearly eliminates the typical 2D cell viability map seen after treatment (Figure 8d) and the larger 10 x 6 well sheet is able to completely counteract the anti-melanoma effect (Figure 8f). This data led us to theorize that the EM waves produced by the CAP jet play an important role in causing the anti-melanoma effect seen after physical CAP treatment. This conclusion also supports the phenomenon that bubbling will still exist even if the tip of the CAP jet doesn't directly touch the back surface of dish (Figure 7).

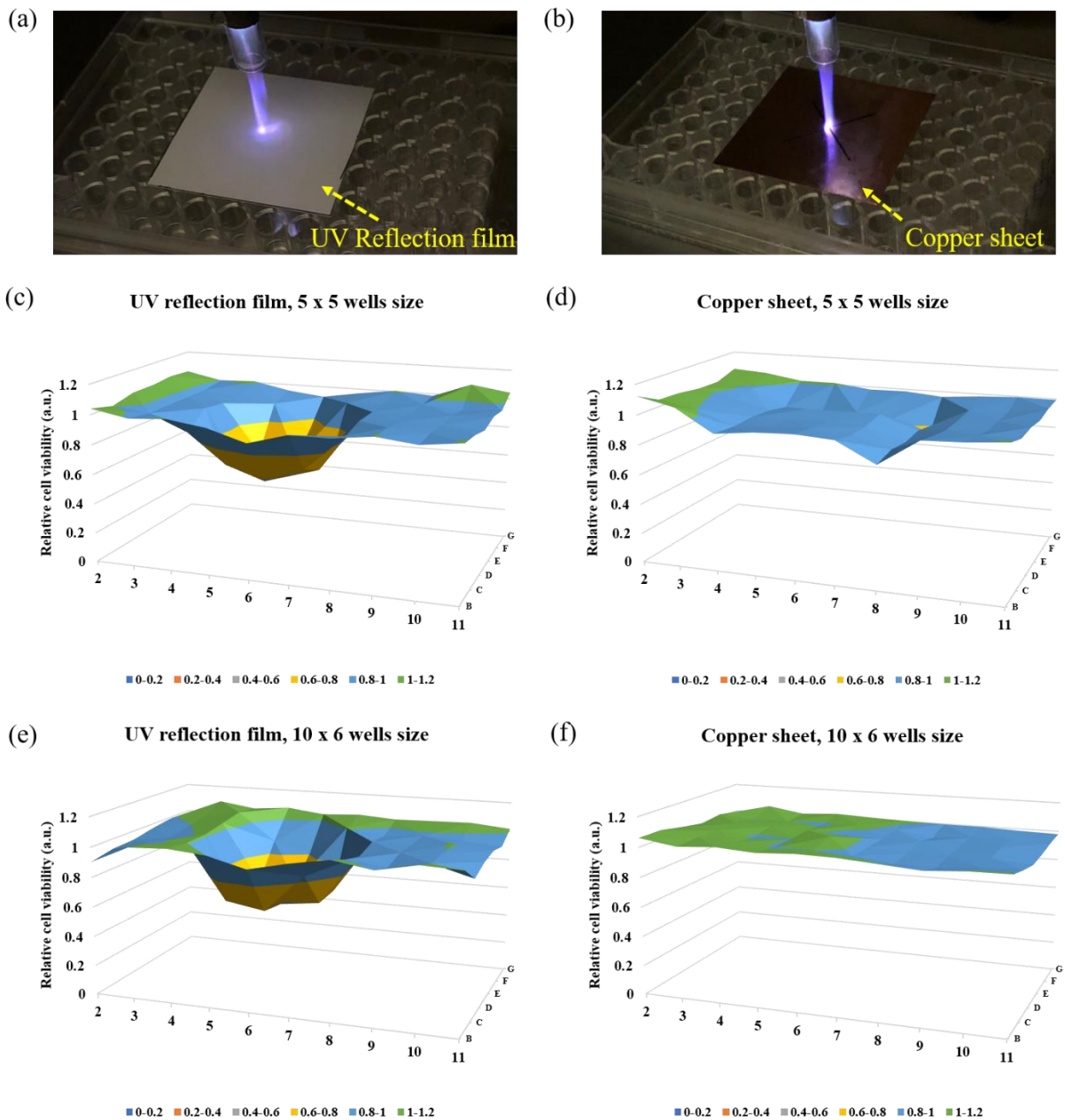

Figure 8. The physically based anti-melanoma effects can only be blocked by a copper sheet, rather than by a UV reflection film. The photos of the physically based CAP treatments on the bottom of 96-well plates covered by UV reflection and copper sheet are shown in a and b, respectively. Here, we just use the UV reflection film or the copper sheet with a size of 5 x 5 wells size as an example. The corresponding 2D cell viability is shown in c to f. The original data were shown in Figure S10. The treatment was 8 minutes. The initial cell density was 6 x $10^4$ cells/mL. The cancer cells

were cultured one day before the final MTT assay. The gap between the nozzle to the bottom surface was 25 mm. The treatment has been independently repeated for 4 times.

Though the chemical components of CAP have been extensively studied over the past decade using optical emission spectrum (OES) and laser-induced fluorescence, the EM wave generation in the Radio Frequency (RF) range by CAP is unknown. A heterodyne setup was designed to measure the RF emission from the CAP jet (Figure 9a). The description of the RF spectrum measurement is illustrated in Methods and Material. Figure 9b shows the RF emission spectra of the CAP jet under different discharge voltages (pk-pk). The spectrum has 4 peaks. For the peaks at around 10 GHz and 14.5 GHz, a higher discharge voltage leads to a lower power density. However, for the two peaks at a higher frequency, increasing the discharge voltage may first increase the power and subsequently decrease it. The power density reaches its maximum value when the discharge voltage is around 8 kV (pk-pk). Besides, for some power density peaks, the frequency shifts can also be observed on the spectrum. This is the first demonstration of the RF range EM waves generated by the CAP jet. This measurement supports the conclusion that the EM waves generated by CAP can trigger the physically based cell death.

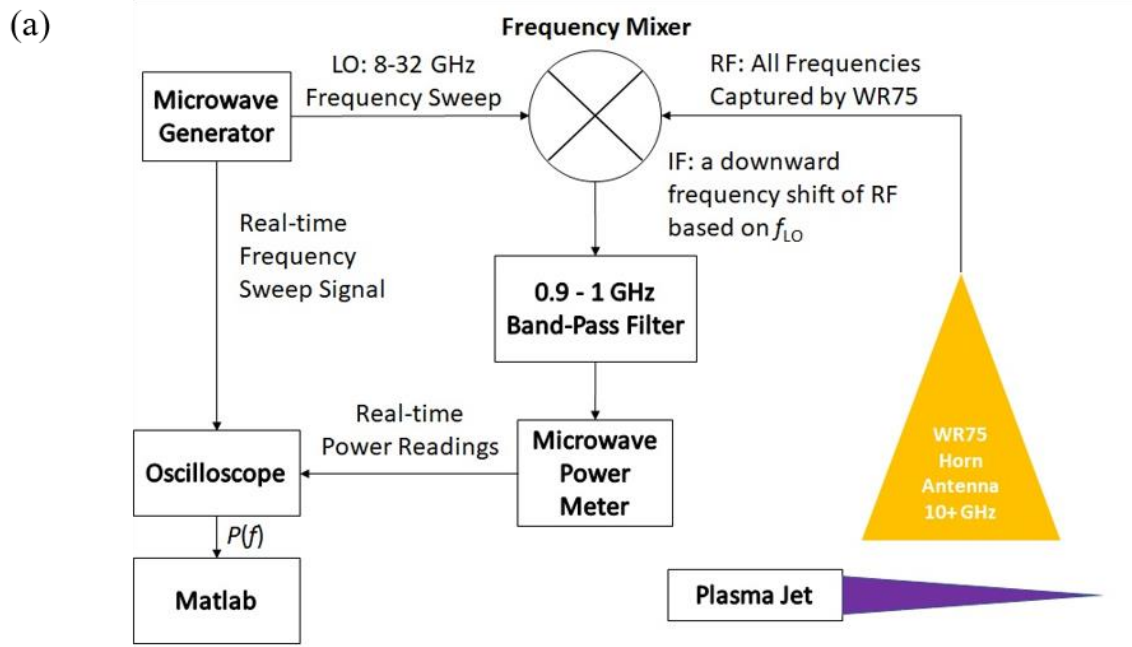

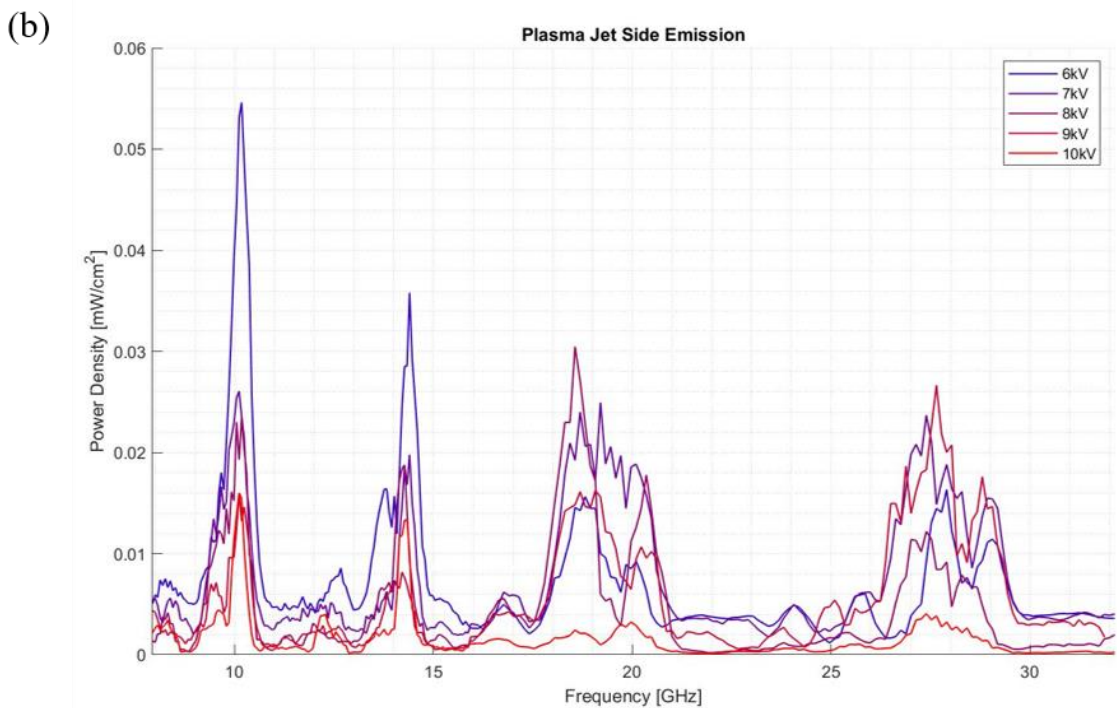

Figure 9. The EM emission of CAP jet under the different discharge voltages (pk-pk). (a) The heterodyne setup for RF power spectrum measurement. (b) The RF spectra of CAP jet over 8-32 GHz.

**The key role of the medium in the direct CAP treatment *in vitro*.**

Over the past decade, the main focus in plasma medicine has been on the dominant role of reactive species in the CAP-cells interaction, particularly the cytotoxicity of CAP on mammalian cells such as cancer cells [18]. In this study, we demonstrated an even stronger anti-cancer effect can be achieved by the physical factors of CAP. We theorize that these effects have not been seen because nearly all prior *in vitro* studies had a medium layer covering the cells. We believe that EM waves generated by CAP treatment can be absorbed by bulk aqueous solution, therefore, only the chemical effect of CAP was seen in the *in vitro* studies done previously. The biological effects seen after physical CAP treatment should be observed when treatment is done on the back surface of the dish or on the cell directly without the media coverage.

To test this hypothesis, we performed a direct CAP treatment on cells without the coverage by the medium as well as a treatment on cells with the varying volume of medium coverage (Figure 10). It is found that direct CAP treatment on exposed B16F10 cells (no medium coverage) resulted in shrinkage of the cytoplasm and bubbling from the cell membrane. The 10 mL and 20 mL of the medium layer in 100 mm dish had a thickness of 1.3 mm and 2.6 mm, respectively. In both cases, 10 mL and 20 mL, the physical effects of CAP were not observed. In other words, just a medium layer 1.3 mm can block the EM waves generated from CAP *in vitro*. This may be the reason that all previous studies could not find evidence of a physical effect if following standard *in vitro* culture conditions. Even the similar cell death has been observed before, it is difficult to investigate the underlying mechanism due to the complex nature of the direct interaction between the CAP jet and the cells.

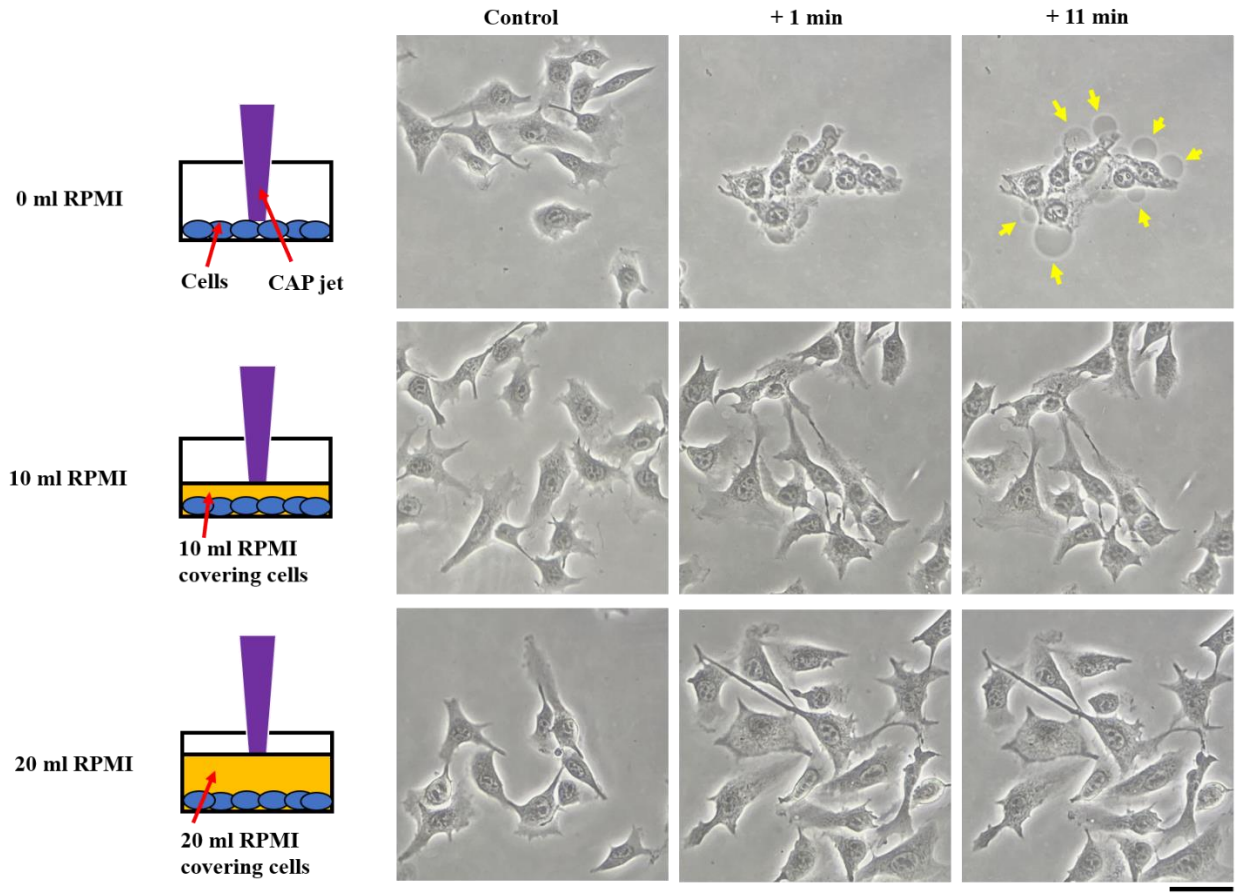

Figure 10. A layer of medium can completely block the physically triggered bubbling. 10 mL cell solution was cultured in 100 mm dish with a density of 7.5 x 10$^4$ cells/mL for one day before the treatment. The flow rate was 1.53 lpm. The gap between the nozzle to the cells was 19 mm. Different from above cases, the CAP treatment at here was performed by directly touching the melanoma cells without or with a layer of medium with different volumes (10 mL and 20 mL). After the treatment, the cells were immediately (< 30 s) immersed in 10 mL of RPMI for the microscopic imaging. Because the initial bubbles were not clear, only the bubbles after a 11 min of growth were marked by yellow arrows. The scale bar was 50 µm (black). The photos of control and the experimental group were taken at different places on the dish. The photos of the experimental were taken in situ. '+ x min' means the photo was taken at x min after the CAP treatment.

We further investigated whether the bubbling could be inhibited by exposure to various hypotonic solutions after direct CAP treatment on the cells without a medium coverage. Similar to the trend shown in Fig. 6, we found an increase in the volume ratio of Milli-Q water gradually inhibited bubble production and the bubble growth are completely inhibited in solely Milli-Q solution (Figure 11). These results indicate that the physical effect of the CAP jet will impact cancer cells in the same way regardless of treatment from the back of a plate or dish or a direct treatment of cells without the medium coverage. We conclude that medium blocks physically based cytotoxicity, therefore, only the chemical factors will play an anti-cancer role in this case as was found in prior studies.

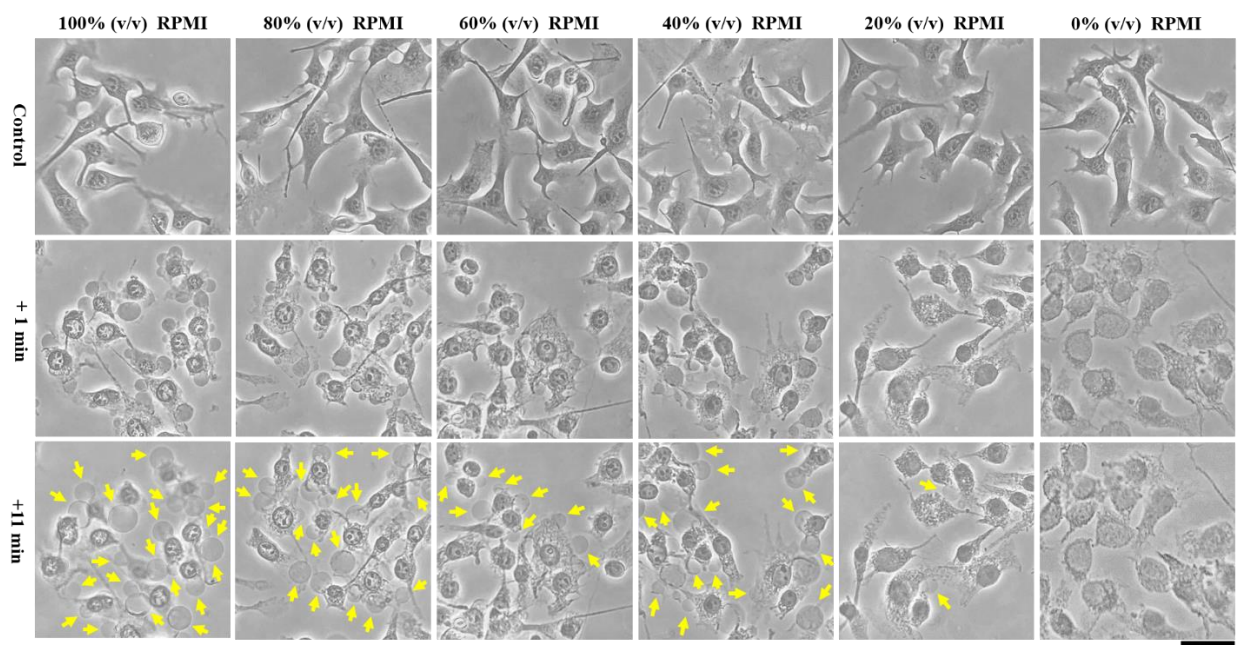

Figure 11. The hypotonic solution can inhibit the bubbling on the direct CAP treatment on melanoma cells without the coverage of a medium layer. A 3 min of CAP treatment was performed on the cells without the coverage of a medium layer. The treated cells were immediately (< 30 s) immersed in 10 mL of RPMI/Milli-Q water mixed solutions. The volume ratio % (v/v) of RPMI in the solutions was 100%, 80%, 60%, 40%, 20%, and 0%, respectively. The seeding cell density was 7.5 x $10^4$ cells/mL. 10 mL cell solution was cultured in 100 mm dish for one day before the treatment. The flow rate was 1.53 lpm. The gap between the nozzle to the cells was 19 mm. Because the initial bubbles were not clear, only the bubbles after a 11 min of growth were marked by yellow arrows. The scale bar was 50 µm (black). The photos of control and the experimental group were taken at different places on the dish. The photos of the experimental were taken in situ. '+ x min' means the photo was taken at x min after the CAP treatment.

**Discussion**

This study is unique because it focused on the physical factors and physical effect of CAP rather than on the previously studied chemical effect and CAP-originated reactive species. This physical factor-based CAP treatment may clarify many of the puzzles in plasma medicine which cannot be explained by the reactive species-based perspective. As illustrated in this study, reactive species are not necessary to achieve a strong anti-cancer effect after CAP treatment. Additionally, the bulk aqueous environment is not a prerequisite for CAP treatment to be effective. It is also known that B16F10 cells are much more sensitive to EM waves than reactive species. In our experiments, we found that physical CAP treatment could affect a larger area of cells compared to chemical CAP treatment. Thus, we provide a solution to overcome the natural limitation of traditional CAP treatment based on the reactive species.

Physical CAP treatment causes a new type of cell death characterized by bulk leakage of water out of the cell and cytosol aggregation (Figure 12). These changes are seen just a couple of minutes after treatment, a much faster process than the typical apoptosis pathway. The bubbling may be due to the formation of hole on the cytoplasmic membrane, which has not been confirmed, by can be speculated from the observation of growth of new bubbles (video 1 and video 2). The long-term changes seen in the cells after physical CAP treatment is also unique. The cells gradually lose DNA from the nucleus over the course of one to two days after treatment. The aggregated cytoskeleton stays intact for 3-5 days after treatment, which may finally form an empty shell. We called such an empty shell as the 'fixed' cells after the treatment. These features are very different from apoptotic features and have never been reported before.

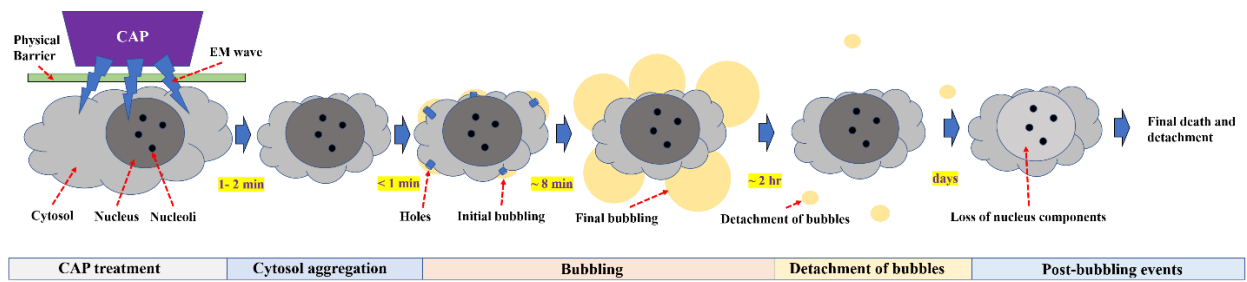

Figure 12. The schematic illustration of the whole process of physically triggered cell death. The cell death has main four stages: cytosol aggregation, bubbling, detachment of bubbles, as well as the post-bubbling events. The first two stages are quick cellular response, totally lasting around 10 min. The detachment of bubbles may last more than 1 -2 hr. In contrast, the post-bubbling events may last days. The bubbling may be a process that the cells drastically lose it water components. The main cytosol shape will not change after the first stage, which gives the cells an appearance as have been 'fixed' after the treatment. Such a 'fix shape' will be kept over days until the final cell death or detachment.

Physical CAP treatment builds a connection between *in vitro* and *in vivo* experiments. Previous studies have concluded that apoptosis is the main form of cell death following chemical CAP treatment. The apoptosis will not trigger an immune or inflammation response *in vitro*, however, several *in vivo* animal studies found a noticeable immune response during the CAP treatment [19–22]. The conventional reactive species-based plasma medicine perspective cannot explain this immunologic dilemma. However, this newly discovered cell death, a clear necrosis involving bulk leakage of water or cellular solutions will likely trigger an immune response *in vivo* if the same changes that occur *in vitro* are seen during *in vivo* studies.

The transdermal diffusion of transportation is necessary for the explanation of the *in vivo* anti-cancer effect of the reactive species-based CAP treatment [23,24]. However, there is no obvious evidence so far indicating the existence of a strong transdermal diffusion of ROS, particularly $H_2O_2$, an important anti-cancer reactive species *in vitro* [25]. As demonstrated in this study, the physical effect can penetrate a dielectric material with a thickness of 1.3 mm, which is close to the thickness of human skin. Our observations suggest that it may not be necessary to assume that transdermal diffusion of reactive species exists when CAP treatment is performed above animal skin.

The underlying mechanism of the physical anti-melanoma effect is still unknown. Based on the microscopic observation, the aggregation may trigger the bubbling by forming an intracellular osmotic pressure. And, the holes may form on the cellular membrane to facilitate the bubbling. The connection between the EM wave and the cytosol aggregation is also unknown. The reason to cause the 'fixed' shape of cells after the treatment is also unknown. At this time, we just discuss possible biological effect of the EM wave based on previous references. In the past, the interaction of mm-wavelength EM with living cells was considered extensively. The importance of this interaction stems from the fact that biological species on Earth are not very well accustomed to this type of interaction. This idea was proposed by Devyatkov who believed that living organisms on Earth are not well adapted to this type of radiation because under natural conditions it is practically absent due to strong absorption by the Earth's atmosphere [26]. This is due to the fact that mm-wavelength is absorbed by water molecules in rotational mode. Several important discoveries were reported such as the dependence of EM action on cells on frequency when various microorganisms are irradiated by mm-waves [27]. Besides, *in vivo* studies were performed to evaluate the effect of mm-wave on tumors and the growth of cells damaged by the ionizing radiation [28]. To that end, the isolated cells damaged by ionizing radiation were consequently irradiated with EM waves at frequencies of 54-76 GHz and power density of $10^{-16}$-$10^{-14}$ W/cm$^2$ for seven minutes [28]. The effect of mm-waves at frequencies of 35.9-55.1 GHz on the growth of implanted tumors (such as carcinomas) in mice and rats has been studied [28]. The power was applied via acupuncture points. These experiments showed that mm-waves at non-thermal intensities act to normalize the growth of cells damaged by the ionizing radiation, and in their action on biologically active zones in animals they have an immunomodulating effect. The CAP jet generates EM waves having a similar frequency as shown in Figure 9, which is likely causing the observed cell death.

**Conclusion**

The physical factors of CAP treatment show strong growth inhibition on B16F10 cells, a typical melanoma cell line. Compared to traditional chemical CAP treatment, physical CAP treatment shows a much stronger growth inhibition on melanoma cells *in vitro*. The electromagnetic waves emitted from the CAP jet could cause cell death through a physical barrier (~1 mm) without contacting the cells as well as through direct contact with cells not covered by medium. Cell death following physical CAP treatment is characterized by rapid bulk leakage of water through the cell membrane accompanied by the cytosol aggregation. The bubbling seen on the cell membrane is a typical feature of necrotic cell death. The bubbling may be triggered by the physically triggered osmotic pressure, which can be inhibited in the hypotonic solutions. This study builds the foundation for using CAP as a non-invasive anti-cancer therapy and possibly applying CAP treatment in various other branches of medicine.

**Methods and Materials**

**CAP jet.**

The CAP jet was designed and assembled at Keidar's lab at the George Washington University. The detailed introduction can be found in previous publications [17,29]. Briefly, the CAP jet was formed through the discharge (3.15 kV, peak value) between a copper ring grounded cathode and a central stainless anode. The ionized gas was flowed out by helium guided in a glass tube with a diameter of 4.5 mm. The maximum tip temperature of the CAP jet during the treatment was around 40°C.

**Cell culture.**

B16F10 cells were donated by Prof. Eduardo Sotomayor at the George Washington University. The cell culture medium was composed of RPMI-1640 (ATCC 30-2001) supplemented with 10% fetal bovine serum (Atlanta Biologicals, S11150) and 1% (v/v) penicillin and streptomycin solution (Life Technologies, 15140122). For the CAP treatment performed on 96-well plate, B16F10 cells were seeded (100 μL/well) in a 96-well plate with a density of 6 x $10^4$ cells/mL. For the CAP treatment performed on a 12-well plate, B16F10 cells were seeded (1 mL/well) in a 12-well plate with a density of 5 x $10^4$ cells/mL. All cells were cultured 24 hr under the standard culture condition (a humidified, 37°C, 5% $CO_2$ environment) before the experiments.

**Chemically based and physically-based CAP treatment.**

For the cases using a 96-well plate, to perform the chemically based CAP treatment, the medium used in the previous culture for 24 hr was removed first. Then, 100 μL/well of fresh RMPI-1640 medium was added to cover all 10 x 6 wells in the middle of 96-well plate. The chemically-based CAP treatment (1 min, 4 min, or 8 min) was performed after this step. After the treatment, the cancer cells were cultured for two days before the cell viability assay. To perform the physical CAP treatment, the medium used in the previous culture for 24 hr was also removed first. Due to the surface tension and the adhesion of water, there was a thin water layer in each well, particularly at the junction between the wall and the bottom in a well. Though there was no bulk medium left to cover the cells during the treatment, the cell viability of cancer cells did not decrease (Figure S2). After the treatment, 100 μL/well RMPI-1640 was added to culture the cells in the middle 10 x 6 wells. The treated cells were cultured two days before the cell viability assay.

For the cases using a 12-well plate, the general strategy to perform the chemical and the physical treatment were unchanged. A single chemical treatment was performed in a single well with a 1 mL of RPMI-1640 medium. A single physical treatment was performed on the back of a single well without medium coverage. After the treatment, 1 mL/well RMPI-1640 was added to culture the cells for 3 days before the cell viability assay. The wells on the 12-well plate were much larger than those of 96-well plates. The treatment on a single well only affected the treated well. Thus, for the 12-well plate case, we just used the traditional single well's cell viability to describe the physically-based anti-melanoma effect.

**Cell viability assay.**

The cell viability assay was performed using MTT assay according to the protocols provided by the manufacturer (Sigma-Aldrich, M2128). The absorbance at 570 nm was read using an H1 microplate reader (Hybrid Technology). For the 96-well plate, because the whole middle 10 x 6 wells on each plate were used to qualify the anti-cancer effect in each case, we used 2D cell viability map to quantify the anti-melanoma effect. This is a novel strategy to show the cell viability, the detailed protocols were illustrated in Figure S1. The 2D cell viability map shows all the cell viability data of 10 x 6 wells simultaneously. For the 12-well plate, the relative cell viability was obtained by the division between the experimental group and the control.

**Live cell fluorescent imaging.**

We seeded 600 μL B16F10 cells (3 x $10^3$ cells/mL) on the 35 mm confocal observation glass-bottom dish and cultured two days before the treatment. After a proper adherence of cells, the morphological changes in the live melanoma cells were assessed by staining microtubule (BioTracker 488 Green Microtubule Cytoskeleton Dye, Sigma-Aldrich) and DNA (Hoechst 33342 Staining Dye Solution, ThermoFisher Scientific). B16F10 cells were stained with both dyes after an overnight culture incubation. The dye working concentration for the microtubule staining was 1 μl in 1mL (0.1%) and the DNA staining was 1 μg/mL in RPMI-1640 media. The cells were incubated for 15 min inside the $CO_2$ incubator before the CAP treatment.

**RF spectrum measurement.**

The RF emission containing multiple frequency components was collected by a WR75 horn antenna which works for 10+ GHz microwave. The frequency mixer shifted the signal to a lower frequency range based on the local oscillator (LO) signal, which was provided by a microwave generator. Since the LO was a frequency sweeping signal from 8 to 32 GHz, the resulting output intermediate frequency (IF) signal was the same spectrum of RF but keeping a downward frequency shifting. The IF then passed through a 0.9-1 GHz band-pass filter. The microwave power meter thus integrated the total power through the filter window. Since the IF kept shifting, a full picture of emission power was obtained. The microwave power meter provided the power as a function of time, while the microwave generator reported the frequency shift as a function of time. A power spectrum was finally obtained.

**Acknowledgments.**

This work was supported by the National Science Foundation, grant 1747760. The authors would like to thank for the instructive training and warm help from Prof. Anastas Popratiloff at the Nanofabrication and Imaging Center of the George Washington University.

**Conflict of interest.**

The authors declare that they have no competing interests.

**References.**

[1] A.J. Miller, M.C. Mihm, *N. Engl. J. Med.* **2006**, *355(1)*, 51-65.

[2] J.F. Thompson, R.A. Scolyer, R.F. Kefford, *Lancet* **2005**, *365(9460)*, 687-701.

[3] M.A. Linares, A. Zakaria, P. Nizran, *Primary care* **2015**, *42(4)*, 645-659.

[4] M.S. Soengas, S.W. Lowe, *Oncogene* **2003**, *22(20)*, 3138-3151.

[5] A.H.S. Peach, *Surg.* **2006**, *24*, 21.

[6] P.D. Kilmer, *Journal. Theory, Pract. Crit.* **2010**, *11*, 369.

[7] A. Fridman, A. Chirokov, A. Gutsol, *J. Phys. D: Appl. Phys.* **2005**, *38(2)*, p.R1.

[8] M. Laroussi, T. Akan, *Plasma Processes Polym.* **2007**, *4(9)*, 777-788.

[9] D.B. Graves, *Plasma Processes Polym.* **2014**, *11(12)*, 1120-1127.

[10] M. Keidar, *Plasma Sources Sci. Technol.* **2015**, *24(3)*, 033001.

[11] M. Keidar, A. Shashurin, O. Volotskova, M. Ann Stepp, P. Srinivasan, A. Sandler, B. Trink, *Phys. Plasmas* **2013**, *20(5)*, 057101.

[12] D. Yan, J.H. Sherman, M. Keidar, *Oncotarget,* **2017**, *8(9)*, 15977.

[13] I. Adamovich, S.D. Baalrud, A. Bogaerts, P.J. Bruggeman, M. Cappelli, V. Colombo, U. Czarnetzki, U. Ebert, J.G. Eden, P. Favia, D.B. Graves, *J. Phys. D: Appl. Phys.* **2017**, *50(32)*, 323001.

[14] M. Keidar, D. Yan, I.I. Beilis, B. Trink, J.H. Sherman, *Trends Biotechnol.* **2018**, *36(6)*, 586-593.

[15] M. Keidar, *Phys. Plasmas* **2018**, *25(8)*, 083504.

[16] D.B. Graves, *IEEE Trans. Radiat. Plasma Med. Sci.* **2018**, *2(6)*, 594-607.

[17] D. Yan, L. Lin, J.H. Sherman, J. Canady, B. Trink, M. Keidar, *IEEE Trans. Radiat. Plasma Med. Sci.* **2018**, *2(6)*, 618-623.

[18] D.B. Graves, *J. Phys. D: Appl. Phys.* **2012**, *45(26)*, 263001.

[19] K. Mizuno, K. Yonetamari, Y. Shirakawa, T. Akiyama, R. Ono, *J. Phys. D: Appl. Phys.* **2017**, *50(12)*, 12LT01.

[20] A. Lin, B. Truong, S. Patel, N. Kaushik, E.H. Choi, G. Fridman, A. Fridman, V. Miller, *Int. J. Mol. Sci.* **2017**, *18(5)*, 966.

[21] A. Lin, Y. Gorbanev, P. Cos, E. Smits, A. Bogaerts, *Clin. Plasma Med* **2018**, *9,* 9.

[22] A. Lin, Y. Gorbanev, J. De Backer, J. Van Loenhout, W. Van Boxem, F. Lemière, P. Cos, S. Dewilde, E. Smits, A. Bogaerts, *Adv. Sci.* **2019**, *6.6*, 1802062.

[23] X. Lu, M. Keidar, M. Laroussi, E. Choi, E.J. Szili, K. Ostrikov, *Mater. Sci. Eng. R Rep.* **2019**, *138*, 36-59.

[24] E.J. Szili, J.S. Oh, H. Fukuhara, R. Bhatia, N. Gaur, C.K. Nguyen, S.H. Hong, S. Ito, K. Ogawa, C. Kawada, T. Shuin, *Plasma Sources Sci. Technol.* **2018**, *27.1*, 014001.

[25] O. Tsuneo, K. Kimura, A. Tsuchida, *Colloids Surf B: Biointerfaces.* **2007**, *56(1-2)*, 201-209.

[26] N.D. Devyatkov, M.B. Golant, O.V. Betskiy, *Moscow. Radio. Commun.* **1991**,*168*.

[27] A.Z. Smolyanskaya, E.A. Gel'vich, M.B. Golant, *Uspekhi sovremennoy biologii.* **1979**,

*87.3,* 381-392.

[28]   L.S. Bundyuk, A.P. Kuz'menko, N.N. Ryabchenko, G.S. Litvinov, *Phys. Alive.* **1994**, *2.1*, 12-25 (1994).

[29]   M. Keidar, I.I. Beilis, *Plasma Engineering, 2nd Edition*, Elsevier-Academic Press, San Diego, CA **2018**.